\documentclass[12pt]{spieman}  
\usepackage{amsmath,amsfonts,amssymb}
\usepackage{graphicx}
\usepackage{setspace}
\usepackage{tocloft}
\usepackage{booktabs}
\usepackage{multirow}
\usepackage{multicol}
\usepackage{makecell}
\usepackage{upgreek}
\usepackage{soul}
\usepackage{textcomp,mathcomp}

\usepackage{lineno}

\title{Radiation Effects on Scientific CMOS Detectors for X-ray Astronomy: \uppercase\expandafter{\romannumeral2}. Total Ionizing Dose Irradiation}

\author[a]{Mengxi Chen}
\author[a,b,*]{Zhixing Ling}
\author[a,b]{Mingjun Liu}
\author[a,b]{Qinyu Wu}
\author[a,b]{Chen Zhang}
\author[c]{Jiaqiang Liu}
\author[b,c]{Zhenlong Zhang}
\author[a,b]{Weimin Yuan}
\author[a,b,d]{Shuang-Nan Zhang}
\affil[a]{National Astronomical Observatories, Chinese Academy of Sciences, Beijing 100101, People's Republic of China}
\affil[b]{School of Astronomy and Space Sciences, University of Chinese Academy of Sciences, Beijing 100049, People's Republic of China}
\affil[c]{National Space Science Center, Chinese Academy of Sciences, Beijing, 100190, People's Republic of China}
\affil[d]{Institute of High Energy Physics, Chinese Academy of Sciences, Beijing, 100049, People's Republic of China}

\cftpagenumbersoff{figure}
\cftpagenumbersoff{table} 
\begin{document} 
\maketitle

\begin{abstract} 
Complementary metal-oxide-semiconductor (CMOS) detectors are a competitive choice for current and upcoming astronomical missions. To understand the performance variations of CMOS detectors in space environment, we investigate the total ionizing dose effects on custom-made large-format X-ray CMOS detectors. Three CMOS detector samples were irradiated with a $^{60}$Co source with a total dose of 70 krad and 105 krad. We test and compare the performance of these detectors before and after irradiation. After irradiation, the dark current increases by roughly 20 $\sim $100 times, and the readout noise increases from 3 e$^-$ to 6 e$^-$. The bias level at 50 ms integration time decreases by 13 to 18 Digital Number (DN) at -30 $^{\circ}$C. The energy resolution increases from $\sim$ 150 eV to $\sim$ 170 eV at 4.5 keV at -30 $^{\circ}$C. The conversion gain of the detectors varies for less than 2\% after the irradiation. Furthermore, there are about 50 pixels whose bias at 50 ms has changed by more than 20 DN after the exposure to the radiation and about 30 $\sim$ 140 pixels whose readout noise has increased by over 20 e$^-$ at -30 $^{\circ}$C at 50 ms integration time. These results demonstrate that the performances of large-format CMOS detectors do not suffer significant degeneration in space environment.
\end{abstract}


{\noindent \footnotesize\textbf{*}Zhixing Ling, \linkable{lingzhixing@nao.cas.cn} }

\begin{spacing}{2}   

\section{Introduction}
\label{sect:intro}  

Charge-coupled devices (CCDs) have been widely used in astronomy for 40 years. In recent years, a novel silicon detector, the complementary metal-oxide-semiconductor (CMOS) detector, with its rapid readout speed, excellent uniformity, medium-sized geometrical surface\cite{SERVOLI2011137}, low building cost, and relatively warm operating temperature, have become a competitive alternative of CCD. The advantages of CMOS technology make them well-suited for demanding radiation environments, including but not limited to medical imaging, electron microscopy, phone cameras, space applications, etc.\cite{2003ARNPS..53..263J,2022JATIS...8b6001R}

Imaging detectors in space are susceptible to severe long-term particle bombardment from various sources, most often protons, electrons, and high-energy photons.\cite{2020JATIS...6a6002B}. Thus, we should understand and minimize the potential damage brought up by the high-energy particles and electromagnetic radiation in space. Generally, for Si-based detectors, such as CCD and CMOS, there are two main types of damage mechanisms: Total Ionization Damage (TID) and Displacement Damage. \cite{2001sccd.book.....J}. The TID usually increases the dark current and the charge transfer efficiency. On the other hand, displacement damage can cause damage to a detector's silicon lattice structure, which might lead to permanent hot pixels or even detector malfunctions\cite{Marbs2006INVESTIGATINGTI}. 
 
With the application of CCDs on various astronomical missions, e.g., Chandra,\cite{2003SPIE.4851...28G} Swift,\cite{2005SSRv..120..165B} and XMM-Newton,\cite{2001A&A...365L..18S} the irradiation effects on CCD have been extensively investigated.\cite{2005SPIE.5898..212O,2003NIMPA.512..386S,2003ITNS...50.2018L,2011A&A...534A..20P} Although CMOS detectors have attracted more and more attention in space astronomy in recent years, e.g., the Lobster Eye Imager for Astronomy (LEIA) \cite{2022ApJ...941L...2Z,2023RAA....23i5007L} and Einstein Probe (EP) mission, \cite{2018SPIE10699E..25Y,2022hxga.book...86Y} only few irradiation studies have been carried out.\cite{2010SPIE.7742E..0ED,2018Senso..18..514W,article1,2020JATIS...6a6002B,2017ITNS...64...27B,1185167,7790848,9076701,2017ChPhB..26k4212M} In addition, the techniques of CMOS detectors develop rapidly and have great diversity. Therefore, investigations on the irradiation damage on CMOS detectors are necessary for space applications.

As a newly developed scientific CMOS, studying the performance of our detector, EP4K, under irradiation is crucial for the EP mission and other upcoming projects. The displacement damage effects on this detector have been studied in Liu et al.\cite{2023JATIS...9d6003L}. In this work, we focus on the result of the TID experiment of the EP4K detector. 

EP4K is a custom-built sCMOS detector optimized for space X-ray astronomy. It serves as the focal plane detector of the Wide-field X-ray Telescope (WXT) onboard the EP mission.\cite{2018SPIE10699E..25Y,2022hxga.book...86Y}  LEIA, which carried one WXT module of EP, including four prototypes of these CMOS detectors, has been flown in-orbit in July 2022. LEIA's result proved that the EP4K CMOS detectors operate normally in low Earth orbit. 12 WXT modules are onboard the EP satellite -- a wide field soft X-ray survey mission -- which was successfully launched in January 2024. EP4K has a 4k $\times$ 4k pixel array, and each pixel has a size of 15 $\times$ 15 $\upmu$m. It has an epitaxial thickness of 10 $\upmu$m, readout noise of less than 5 e$^-$, and power consumption of 1.6 W. The detailed performance of the detector can be found in Wu et al.\cite{2022PASP..134c5006W,2023PASP..135k5002W}
 
The experimental setup is described in Sect.~\ref{sect:exp}. In Sect.~\ref{sect:res&dis}, we present and discuss the degradation and changes in values of the dark current, readout noise, gain, and energy resolution due to the effect of the TID. We compare our results to those from the proton radiation experiment in Liu et al.\cite{2023JATIS...9d6003L}and show our conclusions in Sect.~\ref{sect:dis&con}.

\section{Experimental Setup}
\label{sect:exp}
\begin{figure} 
\begin{center}
\begin{tabular}{c}
\includegraphics[width=6.25in]{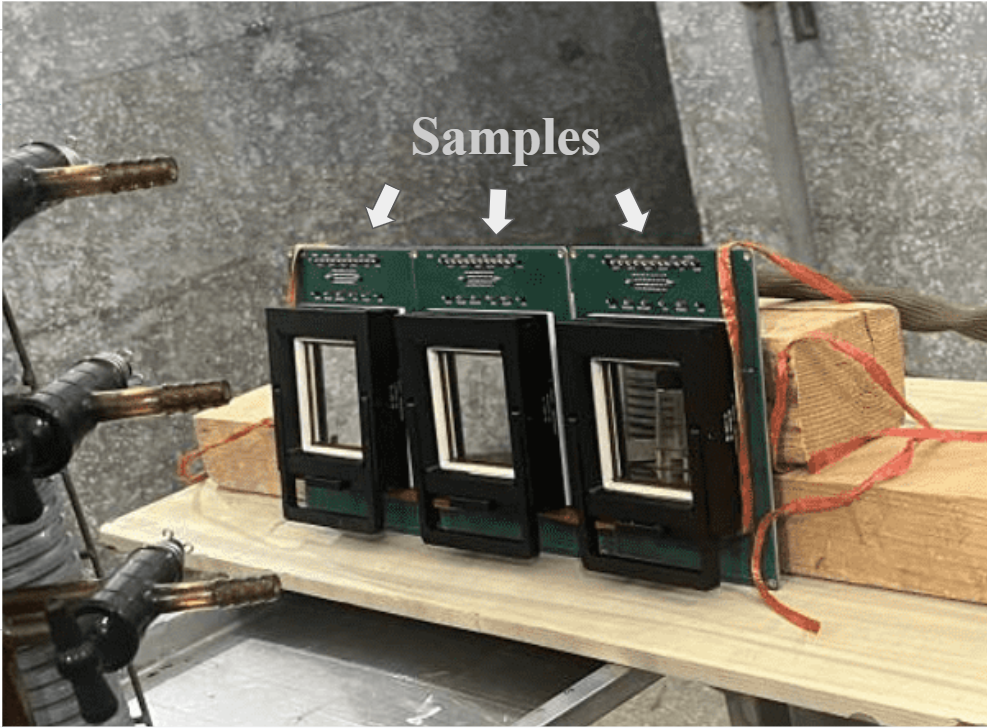}
\end{tabular}
\end{center}
\caption{
\label{fig:equip1}
 A photo of the $^{60}$Co irradiation experimental setup, the three identical sCMOS detector samples are placed on the platform.}
\end{figure}

Radiation experiments of the EP4K CMOS detectors were carried out using the {$^{60}$Co} radiation source at Peking University. The experiment setup is shown in Fig~.\ref{fig:equip1}. Three EP4K CMOS samples were used for the irradiation test. The detector samples we selected in the experiment are from the same batch of flight products for the WXT instrument. After selecting the flight detectors, we randomly chose three available EP4K detectors. The detectors were biased during the irradiation. The simulated TID for the EP in orbit is about 8 krad for its three-year mission duration, and the dominant radiation source is the effective dose of the electrons. To understand the performances of the CMOS detectors during longer operation in orbit, the TID of 70 krad and 105 krad are chosen to evaluate the radiation hardness of the CMOS detector in this work. Therefore, the detectors first went through 70 krad of total radiation dosage under the dose rate of 10 rad$\cdot$s$^{-1}$. After the performance test, the detectors were over-radiated to a total dose of 105 krad and then went through the annealing process in 50 $^{\circ}$C nitrogen atmosphere for 168 hours. Finally, the detectors were tested again. The same camera system \cite{2022SPIE12191E..0LW} was used throughout the experiment. 

\section{Characteristics of performance after the irradiation}
\label{sect:res&dis}

Our analyses are focused on the basic performances and their X-ray spectra properties of the detector samples. The basic performances were tested between -30 $^{\circ}$C to 20 $^{\circ}$C with a interval of 10 $^{\circ}$C. In addition, the performances were also tested at -25 $^{\circ}$C to give a detailed comparison with that of -30 $^{\circ}$C.  Also, the X-ray spectra of the detectors were taken at -30 $^{\circ}$C and 20 $^{\circ}$C.  
We paid particular attention to the operating condition of the EP, in which the CMOS detectors work at a temperature of -30$^{\circ}$C, a readout frame rate of 20 Hz (an integration time of 50 ms). Since each CMOS detector has about $10^2\sim10^3$ inherent defects during fabrication, like that in CCDs, we excluded the bad rows, columns, and pixel clusters in the data processing of this work.

\subsection{Bias Map}
\label{sect:bias}

The bias level is an electronic offset added to the signal from the electronics that ensures that the Analogue-to-Digital Converter (ADC) always receives a positive value. For a CMOS detector, the bias level could be changed by the on-chip register and is varied pixel by pixel. In our analyses, the bias is calculated by averaging the median value of each pixel over 50 images. We analyzed bias maps corresponding to an integration time of 14 $\upmu$s (the shortest exposure time for our detector) and 50 ms (the regular exposure time for EP).

\begin{figure} 
\begin{center}
\begin{tabular}{c}
\includegraphics[width=5.0in]{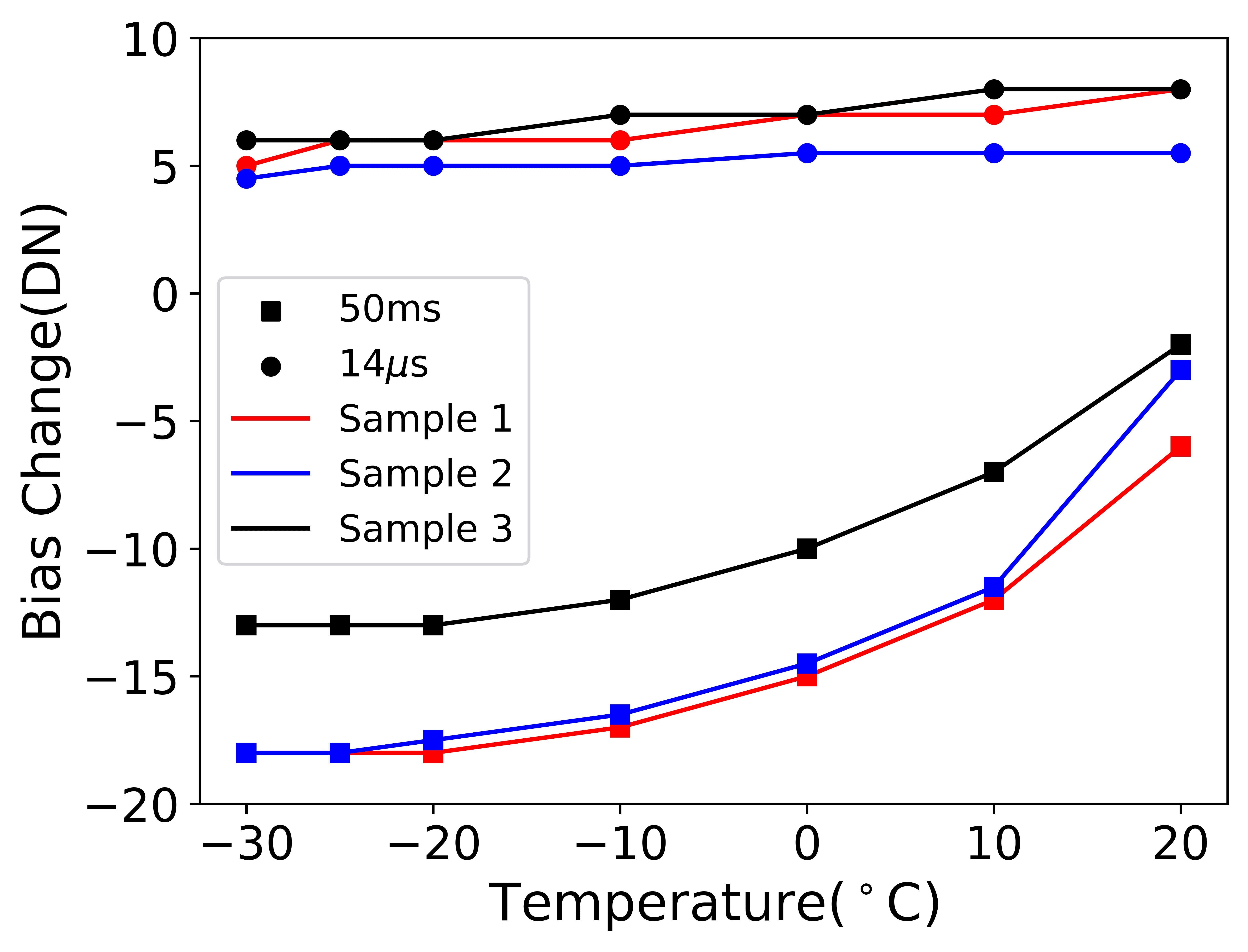}
\end{tabular}
\end{center}
\caption {
\label{fig:change}
The bias change in DN for the bias distribution of all three detectors at 14 $\upmu$s and 50 ms, respectively. The change is calculated as the difference between the median value after 105 krad total dose and the median value before the irradiation.}
\end{figure}

At an exposure time of 14 $\upmu$s, the bias levels of all three detectors have increased by roughly 3 $\sim$ 4 Digital Number (DN) and 5 $\sim$ 6 DN for 70 krad and 105 krad of the total dose separately. At a longer exposure time of 50 ms, bias levels have decreased by 10 $\sim$ 12 DN and 13 $\sim$ 18 DN for 70 krad and 105 krad of total dose among the three detectors. Fig.~\ref{fig:change} shows the bias value changes for all the detectors after exposure to 105 krad total dose (omitted 70 krad data for simplicity) at -30 $^{\circ}$C. The source of the bias change may come from the pixels' pre-amplification or the column readout circuit after the irradiation.

\subsection{Noise}
\begin{figure} 
\begin{center}
\begin{tabular}{c}
\includegraphics[width=6.25in]{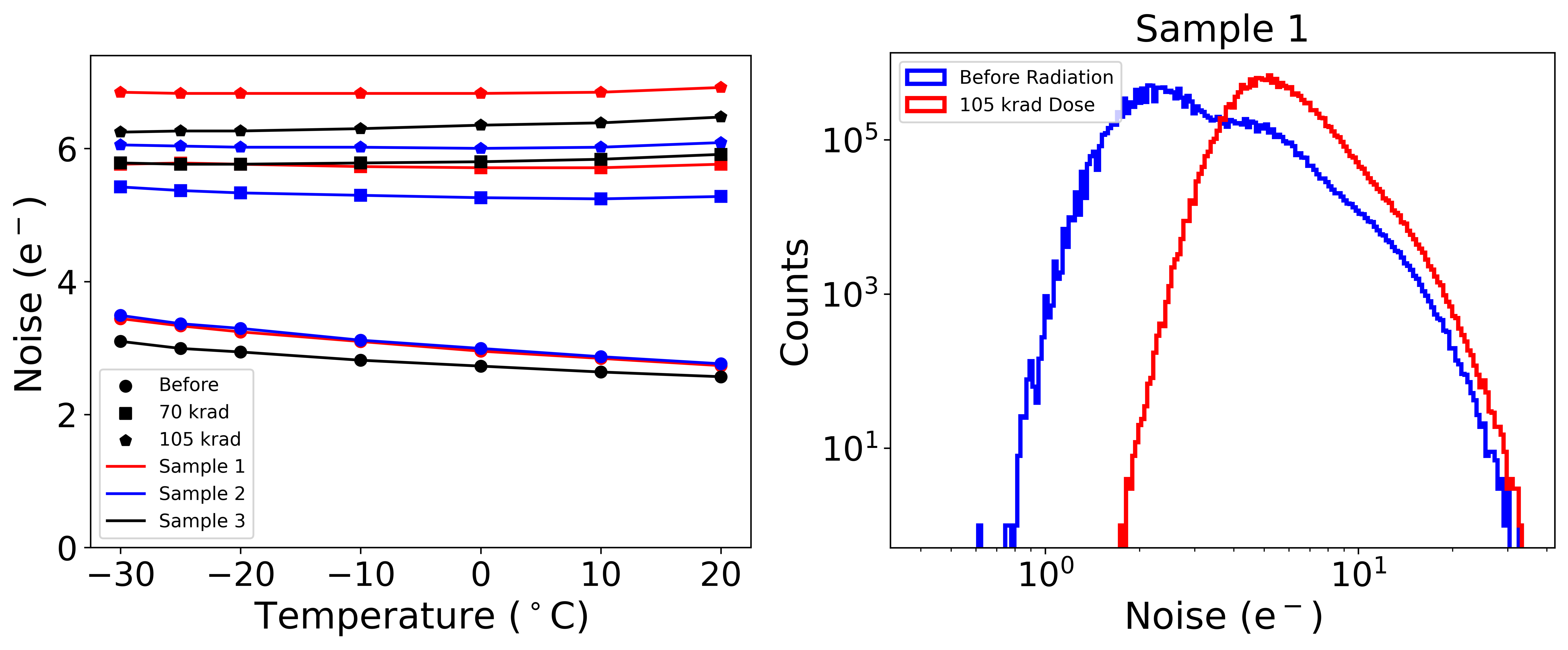}
\end{tabular}
\end{center}
\caption 
{\label{fig:noise}
 The left panel displays each detector's readout noise value (in units of electrons)  before and after irradiation. The right panel shows the distribution of the noise of Sample 1 pixel-by-pixel before and after 105 krad irradiation. The median of the noise distribution has increased for Sample 1; this trend was also found in Samples 2 and 3.} 
\end{figure}

The readout noise of each pixel is calculated as the standard deviation (STD) at the minimum integration time of 14 $\upmu$s over 50 images, while the readout noise of the entire detector is represented by the median of all of the pixels' STD values. The medians of the STD distributions have increased for all three detectors after the irradiation, as seen in Fig.~\ref{fig:noise}. After exposure to the radiation, the noises for all detectors increased from 3 e$^-$ to about 6 e$^-$. All detectors show similar results after the irradiation. 

\subsection{Dark Current}
\begin{figure} 
\begin{center}
\begin{tabular}{c}
\includegraphics[width=5.0in]{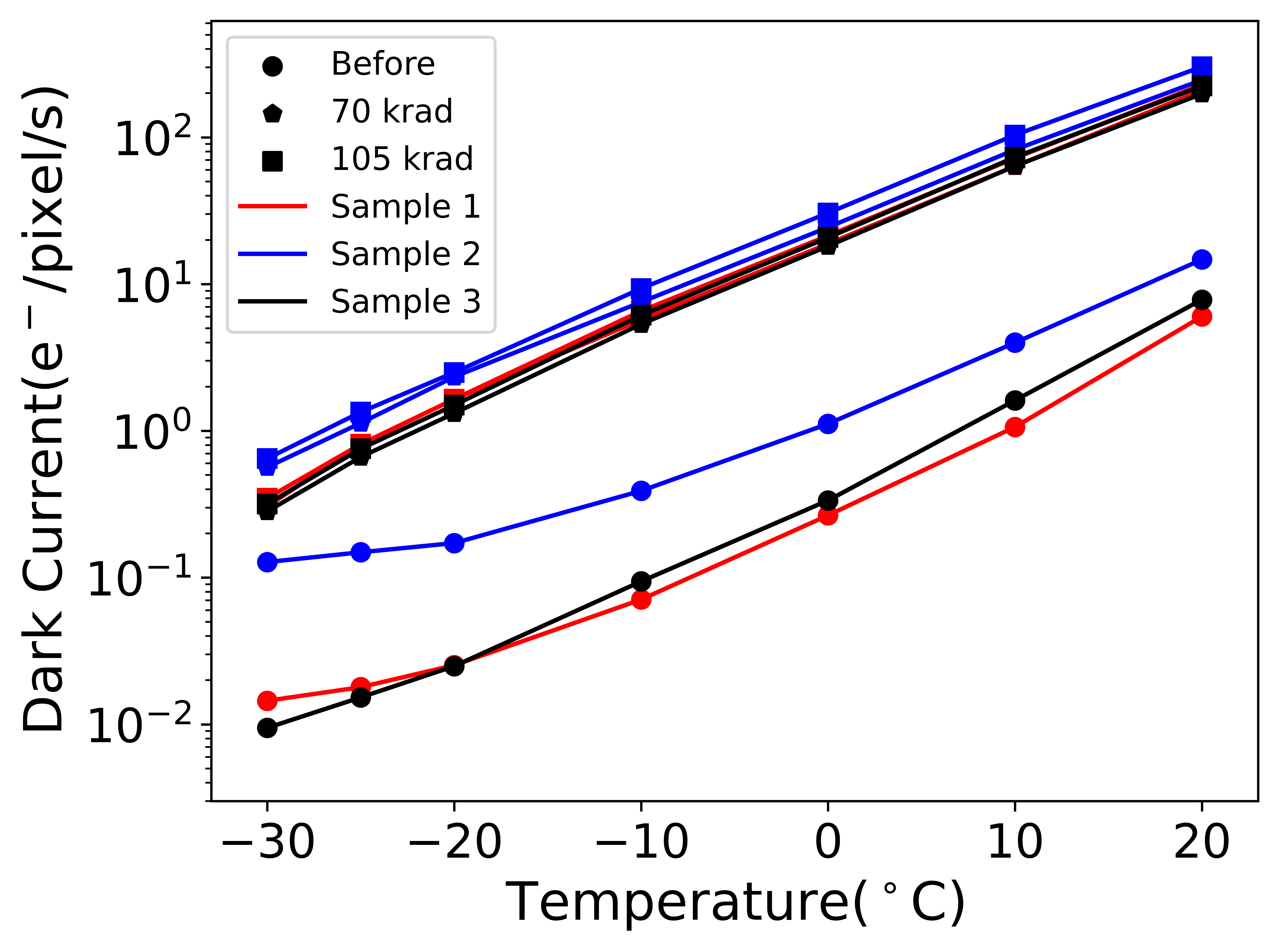}
\end{tabular}
\end{center}
\caption 
{\label{fig:dc}
Dark Current vs. Temperature before and after the irradiation for the three detectors. The three detectors' dark current values behaved similarly before and after the irradiation.)
} 
\end{figure}

The dark current is the self-generated signal of the detector in a dark environment, which is one of the main performance indicators for the CMOS detectors. In this experiment, we calculate the median DN value for all temperatures at each integration time for each frame (averaging over all the available images). Then, the dark current values are calculated at each temperature by linear regression fitting: the slope of the median-integration time plot is the dark current for the CMOS detector at that specific temperature. From Fig.~\ref{fig:dc}, it is clear that after the irradiation, the value of the dark current has increased significantly for all detectors at all temperatures. At -30 $^\circ$C, the dark current increases by more than 20 times. At higher temperatures, the dark current for the three detectors increases by almost 100 times. Although the total dose has a 50\% increase in our experiment (from 70 krad to 105 krad), the corresponding dark current values do not escalate as much. Such phenomenon might have originated from the annealing process, in which the mobile vacancies and defects, produced in the silicon lattice by radiation, reorganize themselves into less harmful and more stable configurations.\cite{article1}

\subsection{Conversion Gain}
The conversion gain of a CMOS detector is the conversion between the number of electrons recorded by the CMOS detector and the number of DN contained in a single image. It is an important parameter for evaluating the performance of the CMOS camera. Tab.~\ref{tab:gain} recorded the conversion gains and the intercepts of the three detectors before and after the exposure to the irradiation at two temperatures. For Samples 1 and 2 at -30 $^{\circ}$C, the conversion gain increases after irradiation, and there is a positive correlation between the total dose and the conversion gain. However, Sample 3's conversion gain dropped after the irradiation (see Tab.~\ref{tab:gain}). The same results are also applicable for each detector at 20 $^{\circ}$C. Even though Sample 3 seems to be an outlier, the variation of the conversion gain is always less than 2\%. Frequent checks of the detectors' conversion gain values in orbit are necessary to calibrate them in space.

\begin{table}[ht]
    \caption{The conversion gains their intercepts of the three sCMOS detectors before and after irradiation.}
    \label{tab:gain}
    \begin{center}
    \begin{tabular}{cccccccc}
    \hline\hline
    \rule[-1ex]{0pt}{3.5ex} \multirow{2}{*}[-1ex]{CMOS} & \multirow{2}{*}[-1ex]{Temperature} & \multicolumn{3}{c}{Gain (eV / DN)} & \multicolumn{3}{c}{Intercept (eV)}\\ \cmidrule(r){3-8}
    \rule[-1ex]{0pt}{3.5ex} & & Before& 70 krad & 100 krad & Before & 70 krad & 100 krad\\
    \hline\hline
    \rule[-1ex]{0pt}{3.5ex} \multirow{2}{*}[-1ex]{Sample 1} & -30 $^{\circ}$C & 6.61$\pm$0.01 & 6.65$\pm$0.05 & 6.69$\pm$0.01 & 41$\pm$5 & 96$\pm$8 & 85$\pm$5\\ \cmidrule(r){2-8}
    \rule[-1ex]{0pt}{3.5ex}  & 20 $^{\circ}$C & 6.51$\pm$0.01 & 6.57$\pm$0.01 & 6.59$\pm$0.02 & 32$\pm$6 & 38$\pm$9 & 158$\pm$10\\
    \hline
    \rule[-1ex]{0pt}{3.5ex} \multirow{2}{*}[-1ex]{Sample 2} & -30 $^{\circ}$C & 6.47$\pm$0.01 & 6.55$\pm$0.01 & 6.57$\pm$0.01 & 33$\pm$3 & 74$\pm$7 & 80$\pm$7\\  \cmidrule(r){2-8}
     \rule[-1ex]{0pt}{3.5ex} & 20 $^{\circ}$C & 6.35$\pm$0.01 & 6.47$\pm$0.01 & 6.47$\pm$0.02 & 27$\pm$5 & 32$\pm$9 & 32$\pm$10\\
     \hline
      \rule[-1ex]{0pt}{3.5ex} \multirow{2}{*}[-1ex]{Sample 3} & -30 $^{\circ}$C & 6.97$\pm$0.01 & 6.74$\pm$0.01 & 6.83$\pm$0.04 & 40$\pm$3 & 83$\pm$9 & 136$\pm$23\\  \cmidrule(r){2-8}
     \rule[-1ex]{0pt}{3.5ex} & 20 $^{\circ}$C & 6.82$\pm$0.01 & 6.69$\pm$0.02 & 6.77$\pm$0.02 & 38$\pm$7 & 56$\pm$11 & 51$\pm$11\\
    \hline\hline
    \end{tabular}
    \end{center}
\end{table}

\subsection{Energy Resolution}
\begin{figure} 
\begin{center}
\begin{tabular}{c}
\includegraphics[width=6.25in]{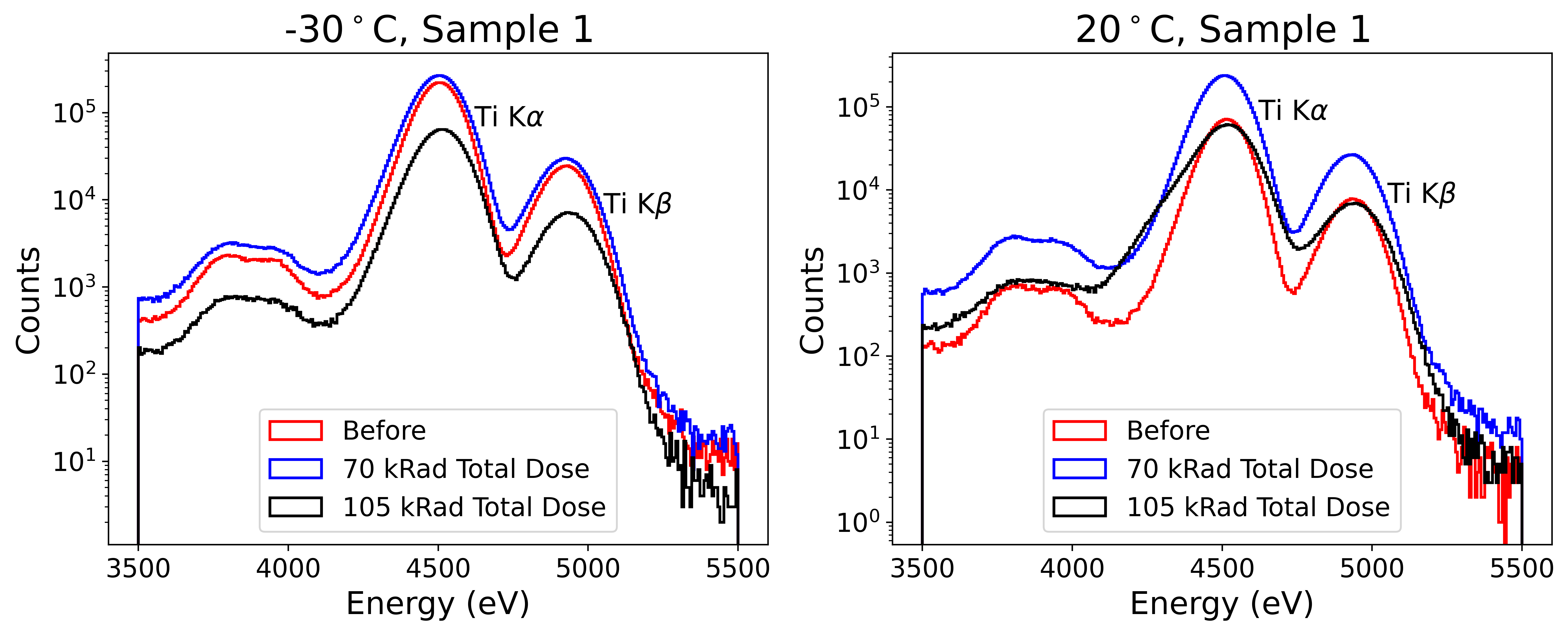}
\end{tabular}
\end{center}
\caption 
{\label{fig:FWHM}
Sample 1 emission line spectra for titanium with 50 ms integration time at -30 $^{\circ}$C (left) and 20 $^{\circ}$C (right). }
\end{figure}

The energy resolution evaluates the ability of one detector to distinguish between distinct emission lines. The full width at half maximum (FWHM) of an emission line is calculated by fitting the emission line in the spectrum using a Gaussian function. The FWHM and their uncertainties of the two distinct energy peaks before and after irradiation are summarized in Tab.~\ref{tab:FWHM}. The spectra of Ti K$_{\alpha}$ and Ti K$_{\beta}$ line of Sample 1 are given in Fig.~\ref{fig:FWHM}. According to Fig.~\ref{fig:FWHM} and Tab.~\ref{tab:FWHM}, at -30 $^{\circ}$C (normal working temperature in space), the FWHM of the Ti K$_{\alpha}$ peak for the three detectors have all increased from 150 eV to about 170 eV, thus suffer moderate degradation. At 20 $^{\circ}$C, Sample 1 suffers significant degradation on Ti K$_{\alpha}$ FWHM (an increase of 47 eV) of the Ti energy peaks, compared to Sample 2 and 3 ($\sim$ 18 eV growth). The main reason for the degradation of energy resolution is the increase in the readout noise. The dark current can also be detrimental to the energy resolution. However, the tiny dark current values for our samples indicate that their effect on the energy resolution is trivial.

\begin{table}[ht]
    \caption{
    \label{tab:FWHM}The energy resolutions (FWHM, in units of eV) of the three CMOS detectors before and after irradiation. The Mg FWHM data for Sample 2 at 20 $^{\circ}$C are not collected.}
    \footnotesize
    \begin{minipage}{.5\linewidth}
    \centering
    \begin{tabular}{lccc}
    \hline\hline
    \rule[-1ex]{0pt}{3.5ex} \multirow{2}{*}[-1ex]{-30 $^{\circ}$C} & \multicolumn{3}{c}{Sample 1} \\ \cmidrule(r){2-4}
    \rule[-1ex]{0pt}{3.0ex} & Before & 70 krad & 100 krad \\
    \hline
    \rule[-1ex]{0pt}{3.5ex}  Ti K$_{\alpha}$ & 155.3$\pm$1.3 & 171.2$\pm$1.3 & 177.1$\pm$1.1\\
    \hline
    \rule[-1ex]{0pt}{3.5ex}  Ti K$_{\beta}$ & 167.9$\pm$1.8 & 183.4$\pm$2.1 & 191.8$\pm$2.2 \\
    \hline
    \rule[-1ex]{0pt}{3.5ex}  Mg K$_{\alpha}$ & 94.2$\pm$3.4 & 114.2$\pm$3.1& 121.6$\pm$3.1 \\
    \hline\hline
    \rule[-1ex]{0pt}{3.5ex} \multirow{2}{*}[-1ex]{-30 $^{\circ}$C} & \multicolumn{3}{c}{Sample 2} \\ \cmidrule(r){2-4}
    \rule[-1ex]{0pt}{3.0ex} & Before & 70 krad& 100 krad \\
    \hline
    \rule[-1ex]{0pt}{3.5ex}  Ti K$_{\alpha}$ & 151.0$\pm$0.9  & 164.4$\pm$1.0 & 168.9$\pm$1.0\\
    \hline
    \rule[-1ex]{0pt}{3.5ex}  Ti K$_{\beta}$ & 163.3$\pm$1.4 & 177.6$\pm$1.7 & 182.1$\pm$1.7\\
    \hline
    \rule[-1ex]{0pt}{3.5ex}  Mg K$_{\alpha}$ & 89.3$\pm$2.5 & 105.1$\pm$2.5& 110.7$\pm$2.5\\
    \hline\hline
    \rule[-1ex]{0pt}{3.5ex} \multirow{2}{*}[-1ex]{-30 $^{\circ}$C} & \multicolumn{3}{c}{Sample 3} \\ \cmidrule(r){2-4}
    \rule[-1ex]{0pt}{3.0ex} & Before & 70 krad & 100 krad \\
    \hline
    \rule[-1ex]{0pt}{3.5ex}  Ti K$_{\alpha}$ & 151.5$\pm$1.6 & 170.1$\pm$1.6 & 176.3$\pm$1.5\\
    \hline
    \rule[-1ex]{0pt}{3.5ex}  Ti K$_{\beta}$ & 164.3$\pm$2.3 & 182.9$\pm$2.2 & 190.0$\pm$2.7 \\
    \hline
    \rule[-1ex]{0pt}{3.5ex}  Mg K$_{\alpha}$ & 96.5$\pm$4.6 & 115.6$\pm$3.4& 121.5$\pm$3.4 \\
    \hline\hline
 
    \end{tabular}
    \end{minipage}%
    \begin{minipage}{.5\linewidth}
    \centering
    \begin{tabular}{lccc}
    \hline\hline
    \rule[-1ex]{0pt}{3.5ex} \multirow{2}{*}[-1ex]{20 $^{\circ}$C} & \multicolumn{3}{c}{Sample 1} \\ \cmidrule(r){2-4}
    \rule[-1ex]{0pt}{3.0ex} & Before & 70 krad & 100 krad \\
    \hline
    \rule[-1ex]{0pt}{3.5ex}  Ti K$_{\alpha}$ & 148.6$\pm$1.3 & 162.8$\pm$0.9 & 195.6$\pm$3.2 \\
    \hline
    \rule[-1ex]{0pt}{3.5ex}  Ti K$_{\beta}$ & 160.3$\pm$1.2 & 175.2$\pm$1.6 & 202.9$\pm$3.4 \\
    \hline
    \rule[-1ex]{0pt}{3.5ex}  Mg K$_{\alpha}$ & 91.2$\pm$4.0 & 111.3$\pm$2.1 & 161.1$\pm$7.0 \\
    \hline\hline
    \rule[-1ex]{0pt}{3.5ex} \multirow{2}{*}[-1ex]{20 $^{\circ}$C} & \multicolumn{3}{c}{Sample 2} \\ \cmidrule(r){2-4}
    \rule[-1ex]{0pt}{3.0ex} & Before & 70 krad & 100 krad \\
    \hline
    \rule[-1ex]{0pt}{3.5ex}  Ti K$_{\alpha}$ & 144.6$\pm$0.9 & 157.2$\pm$0.8 & 162.2$\pm$0.8\\
    \hline
    \rule[-1ex]{0pt}{3.5ex}  Ti K$_{\beta}$ & 156.6$\pm$1.5 & 169.7$\pm$1.4 & 174.0$\pm$1.5\\
    \hline
    \rule[-1ex]{0pt}{3.5ex}  Mg K$_{\alpha}$ &  ------ & ------ & ------\\
    \hline\hline
    \rule[-1ex]{0pt}{3.5ex} \multirow{2}{*}[-1ex]{20 $^{\circ}$C} & \multicolumn{3}{c}{Sample 3} \\ \cmidrule(r){2-4}
    \rule[-1ex]{0pt}{3.0ex} & Before & 70 krad & 100 krad \\
    \hline
    \rule[-1ex]{0pt}{3.5ex}  Ti K$_{\alpha}$ & 147.2$\pm$1.8 & 161.5$\pm$1.4 & 166.5$\pm$1.2\\
    \hline
    \rule[-1ex]{0pt}{3.5ex}  Ti K$_{\beta}$ & 158.5$\pm$2.4 & 173.3$\pm$1.9 & 178.2$\pm$1.9\\
    \hline
    \rule[-1ex]{0pt}{3.5ex}  Mg K$_{\alpha}$ & 94.3$\pm$5.3 & 114.3$\pm$3.0 & 120.0$\pm$2.7\\
    \hline\hline 
    \end{tabular}
    \end{minipage} 
\end{table}

\subsection{Degraded Pixels}
\begin{figure}
\begin{center}
\begin{tabular}{c}
\includegraphics[width=6.25in]{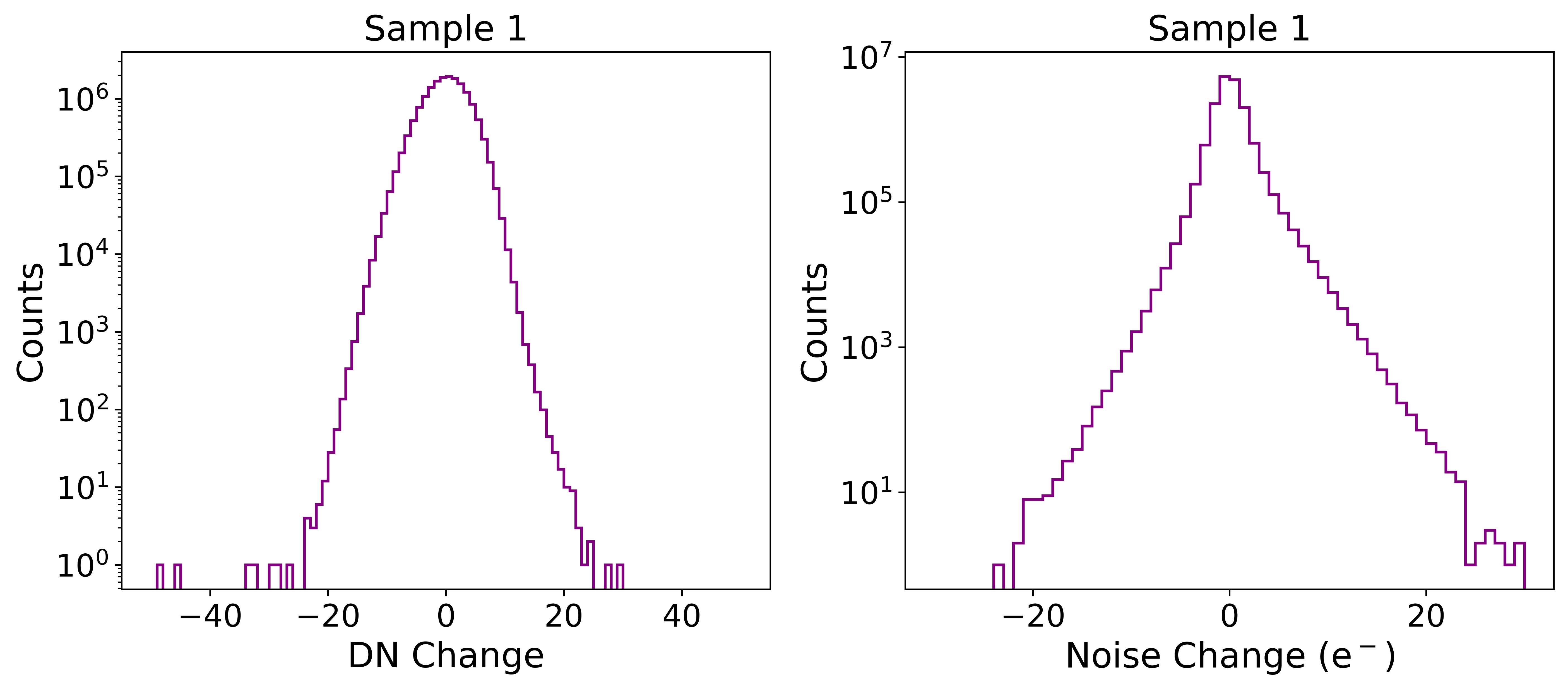}
\end{tabular}
\end{center}
\caption 
{\label{fig:pixel_hist}
Histogram of Sample 1 pixel-by-pixel DN change (left panel) and noise change (right panel) distribution after exposure to 105 krad total dose. There are no pixels that have changes that are beyond the two histograms' range.
} 
\end{figure}

After the irradiation, pixels that had significant changes (in bias, dark current, or noise) were investigated. First, an analysis of the pixel-by-pixel bias change of our detectors after the irradiation at 50 ms integration time is conducted. From Fig.~\ref{fig:pixel_hist}, it is clear that for Sample 1, the histogram of the bias change at the pixel level resembles a Gaussian distribution. A smooth distribution curve indicates that none of the rows, columns, or groups of pixels malfunctioned after the irradiation. The same distribution characteristics can apply to all detectors. The degraded pixels are defined as pixels, at -30 $^{\circ}$C, with a DN change greater than 20 DN after the exposure to 105 krad of total dose or with a change in the noise of more than 20 e$^-$.

\begin{figure}
\begin{center}
\begin{tabular}{c}
\includegraphics[width=6.25in]{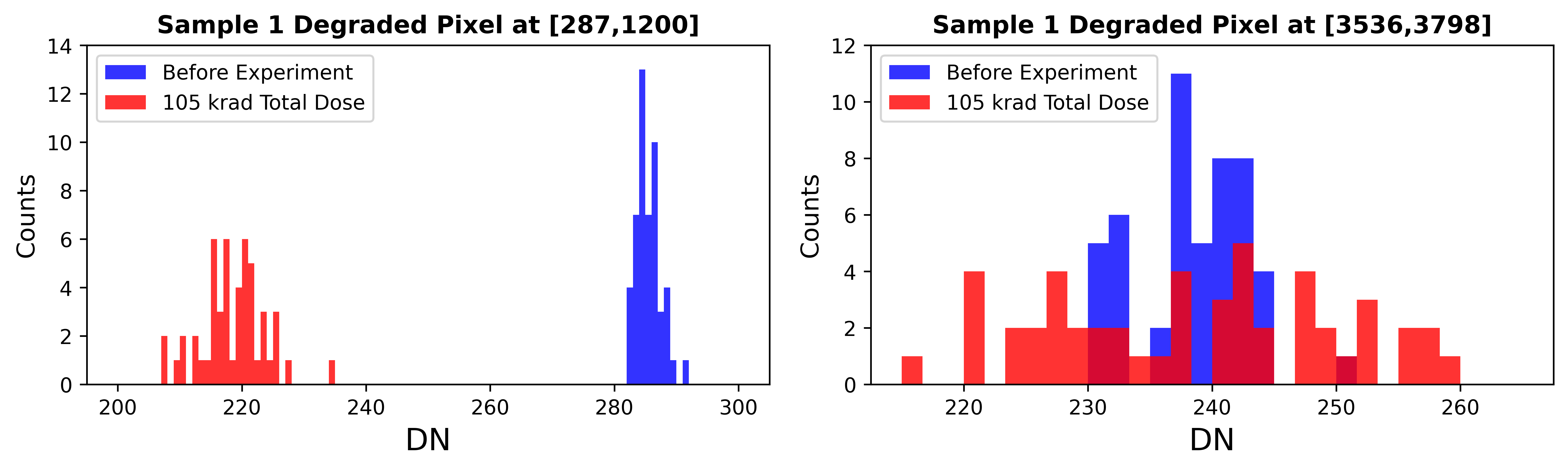}
\end{tabular}
\end{center}
\caption 
{\label{fig:Degraded}
Sample 1 degraded pixels histogram (data of 50 images) with decreasing DN after exposure to the radiation (left panel) and increasing noise (right panel) after exposure at 50 ms integration time.
} 
\end{figure}

Fig.~\ref{fig:Degraded} demonstrates two distinct degraded pixels with significant DN decrease or noise increase after the irradiation. Among all the degraded pixels, no spacial patterns were found on the three detectors, indicating no structural damage on our detectors after the irradiation. Tab.~\ref{tab:count} summarized the total number of degraded pixels in three detectors. Most of the degraded pixels with significant DN change have a decrease in DN, and for the degraded pixels with significant noise change, most of those pixels have some degrees of increase in noise. Since the total number of degraded pixels is relatively small, the degraded pixels can be masked during the detector's in-orbit operation, thus not undermining the performance of EP4K. 

\begin{table}
    \caption{CMOS degraded pixel counts. The "DN" column represents the number of pixels in a detector that has DN distribution median change greater than 20 DN, and the "Noise" column represents the pixel where its noise changed greater than 20 e$^-$ after exposure to 105 krad of total radiation dose, at 50 ms integration time. }
    \label{tab:count}
    \begin{center}
    \begin{tabular}{lcccc}
    \hline\hline
    \rule[-1ex]{0pt}{3.5ex}  Sample & DN & Noise & Total 
    \\
    \hline
    \rule[-1ex]{0pt}{3.5ex}  1 & 83 & 140 & 223 \\
    \hline
    \rule[-1ex]{0pt}{3.5ex}  2 & 43  & 30  & 73 \\
    \hline
    \rule[-1ex]{0pt}{3.5ex}  3 & 51 & 104 & 155  \\
    \hline\hline
    \end{tabular}
    \end{center}
\end{table}

\section{Discussion and conclusion}
\label{sect:dis&con}

In this work, we investigate the total ionization damage and the displacement damage effects on CMOS detectors. According to our previous experiment\cite{2023JATIS...9d6003L}, the displacement damage from the proton irradiation has impacted EP4K differently. The displacement damage for sCMOS mainly leads to increased dark current. After 50 MeV proton irradiation with 8.4 krad of the total ionizing dose, the dark current increases by 110 to 447 times after the irradiation, varying with temperature; however, the readout noise, bias map, gain, and the energy resolution at -30 $^\circ$C hardly changed. Compared to displacement damage, the TID damage has more noticeable effects on the bias map, readout noise, dark current, and energy resolution.

After irradiating with 105 krad total dose at 14 $\upmu$s exposure time, the bias values increase by 5 DN for all detectors at -30 $^\circ$C. At 50 ms exposure times, the bias levels decrease by 13 $\sim$ 18 DN at -30 $^\circ$C for three detectors. Similar bias properties can be observed at the irradiation dosage of 70 krad. For all samples, the readout noise increases by 3 e$^-$ after the irradiation, and the dark current increases by 20 $\sim$ 100 times for both irradiation dosage, depending on the temperature. The conversion gain has less than 2\% fluctuation at both -30 $^\circ$C and 20 $^\circ$C for all detectors. The energy resolution suffers moderate degradation at -30 $^\circ$C for the three detectors. At 20 $^\circ$C, Sample 1 suffers severe degradation, with an increase of 47 eV for the FWHM of the Ti peak. At the same time, Samples 2 and 3 seem to be less affected by the temperature change since the same FWHM peak only has an increase of 18 eV after the irradiation. For degraded pixels, 30 $\sim$ 140 pixels have variations of more than 20 DN in terms of bias and 20 e$^-$ in terms of noise at 50 ms. 
Fortunately, these TID damages are negligible during the EP mission. The changes on the bias map, noise, dark current, and energy resolution are insignificant at -30 $^\circ$C. The fluctuations in conversion gain can be adjusted through regular checks. The number of degraded pixels is manageable since it is well below the design limit for the WXT's front electronics, and the degraded pixels can be masked out by routine scrutiny using dark exposures. The irradiated dose that we used in the experiment is well over the expected total dose for the EP mission and our samples only suffered moderately. The large-format sCMOS detector, EP4K, has a robust performance under the harsh TID radiation environment in low Earth orbit, and is promising for future X-ray missions.

\section* {Code, Data, and Materials Availability} 
The code and data underlying this article will be shared on reasonable request
to the corresponding author.

\section* {Acknowledgments}
This work is supported by the National Natural Science Foundation of China (grant No. 12173055, 12333007 and 12027803) and the Chinese Academy of Sciences (grant Nos. XDA15310000, XDA15310100, XDA15310300, XDA15052100).


\bibliographystyle{spiejour}   
\bibliography{ref}   




\vspace{1ex}
\noindent Biographies and photographs of the authors are not available.

\end{spacing}
\end{document}